\documentclass[aps,prl,twocolumn,groupedaddress,showpacs]{revtex4}
\usepackage{graphicx}
\usepackage{dcolumn}
\usepackage{bm}
\begin{document}

\preprint{APS/123-QED}

\title{ Flow Phase Diagram for the Helium Superfluids }

\author{L. Skrbek }

\affiliation{ Joint Low Temperature Laboratory, Institute of
Physics ASCR and Charles University, \\V
  Hole\v{s}ovi\v{c}k\'ach 2, 180\,00 Prague, Czech Republic}%

\date{\today}
\begin{abstract}
The flow phase diagram for He II and $^3$He-B is established and
discussed based on available experimental data and the theory of
Volovik [JETP Letters {\bf{78}} (2003) 553].  The effective
temperature - dependent but scale - independent Reynolds number
$Re_{eff}=1/q=(1+\alpha')/\alpha$, where $\alpha$ and $\alpha'$
are the mutual friction parameters and the superfluid Reynolds
number characterizing the circulation of the superfluid component
in units of the circulation quantum are used as the dynamic
parameters. In particular, the flow diagram allows identification
of experimentally observed turbulent states I and II in
counterflowing He II with the turbulent regimes suggested by
Volovik.

\end{abstract}

\pacs{ 67.40.Vs, 67.57.De, 47.27.Ak
}
\maketitle

We consider the flow of quantum liquids such as He~II or $^3$He-B
that can be described in the framework of the two fluid model
(see, e.g., \cite{Tough}). The normal fluid and superfluid
velocity fields are coupled by two terms - by the Gorter- Mellink
term that describes the mutual friction between these two liquids
when vortices are present in the superfluid, and by the
temperature gradient term, responsible, e.g., for the fountain
effect. Circulation in the superfluid component is quantized in
units of $\kappa$ ($0.997\times10^{-3}$~cm$^2$/s for He II and
$0.662\times10^{-3}$~cm$^2$/s for $^3$He-B); we assume singly
quantized vortices.

Let us consider a flow that can be approximated as isothermal
\cite{isothermal}. Then the generally coupled complex flow of both
components described by the two two-fluid equations can be
simplified and becomes easier to understand, especially for two
extreme cases:

i) {\it{There are no quantized vortices in the flow.}} This
represents a situation when the normal and superfluid velocity
fields are fully decoupled. The normal fluid thus obeys the usual
Navier-Stokes equation (and standard fluid dynamics) while the
superfluid flow stays potential. Thus formally the normal fluid
could become turbulent without a single vortex being present in
the superfluid - in the absence of mutual friction the superfluid
simply does not know what is happening in the normal fluid. In
practice, however, remnant vortices are almost always present, at
least in He II \cite{Awschalom}, pinned to walls which are always
rough on the atomic scale. In $^{3}$He-B a vortex free sample is
more likely, but the highly viscous normal fluid can hardly become
turbulent in the laboratory sized container.

ii) {\it{The normal fluid is at rest in some frame of reference.}}
Such a possibility arises, for example, for $^3$He-B, whose highly
viscous normal component is effectively clamped by the walls in a
laboratory size container. Moreover, we assume that the quantized
vortices in the flow are arranged in such a way that the
coarse-grained hydrodynamic equation
\begin{equation}
\frac{\partial \textbf{v}_{s}}{\partial t}+ {\boldsymbol\nabla}
\mu =(1-\alpha
')\textbf{v}_{s}\times{\boldsymbol\omega}+\alpha\hat{{\boldsymbol\omega}}\times({\boldsymbol\omega}
\times \textbf{v}_{s})
\end{equation}
obtained from the Euler equation after averaging over vortex lines
\cite{Sonin}, written in the frame of reference of the normal
fluid, provides a sufficiently accurate description of the
superflow. We shall return to the applicability of this equation
later. The normal fluid thus provides a unique frame of reference
and we have to deal only with the superfluid velocity
$\mathbf{v}$. By re-scaling the time such that
$\tilde{t}=(1-\alpha' t)$, when dropping the tilde sign, one gets

\begin{equation}
\frac{\partial \textbf{v}}{\partial t}+ {\boldsymbol\nabla} \mu
=\textbf{v}\times{\boldsymbol\omega}+q\hat{{\boldsymbol\omega}}\times({\boldsymbol\omega}\times
\mathbf{v})
\end{equation}
The theoretical analysis of the fluid dynamical problem based on
this equation has been performed by Volovik \cite{GrishaJETP}. As
was first emphasized by Finne {\it {et al}}  \cite{N}, it has a
very remarkable property which makes it distinct from the ordinary
Navier-Stokes equation where the relative importance of the
inertial and dissipative terms is given by the Reynolds number,
which in turn depends on the geometry of the particular flow under
study. Here the role of the effective Reynolds number is played by
the parameter $Re_{eff}=q^{-1}=(1+\alpha')/\alpha$ that depends on
temperature but not on geometry. We stress that the superfluid
Reynolds number is not relevant to consideration of the problem of
flow obeying eq.(2), the beauty of which consists in the fact that
one is able to derive more general conclusions about turbulent
flow generated from suitable initial conditions depending only on
a single temperature dependent parameter $1/q$, regardless of the
actual geometry of the flow. A wide range of $q$ values is easily
experimentally achievable; with $q$ increasing with temperature in
both He II \cite{Russ} and $^{3}$He-B \cite{Bevan}.

Like the usual Navier-Stokes equation, eq.(2) is a non-linear
differential equation allowing for both stable and unstable
solutions including fully developed turbulence. For $q\gg 1$,
similar to the low Re classical fluid dynamics, the solutions are
stable. As $q$ approaches unity, solutions become unstable and for
$q \ll 1$ it describes fully developed turbulence. The latter is
discussed in detail in \cite{GrishaJETP}, showing that a turbulent
cascade will develop, covering scales from the outer scale of the
order of the container size, $R$, down to a minimum scale
$r_{o}\sim q^{3/2}R$ with the velocity scale $v_{r0} \sim
q^{1/2}U$, $U$ being the velocity at the outer scale $R$. The 3D
energy spectrum remains in its usual Kolmogorov form $E(k)\cong
\varepsilon^{2/3} k^{-5/3}$ and the usual relation for the energy
decay rate $\varepsilon=-dE/dt=v_{r}^{3}/r = U^3/R$ also holds in
the $q \ll 1$ limit. The dissipation mechanism is, however,
different in that instead $\varepsilon=\nu \omega^2$ for classical
turbulence it now depends explicitly on the large scale velocity:
$\varepsilon=-dE/dt\sim q\omega U^2$.

It is even possible \cite{GrishaNote} to work out logarithmic
corrections to this approach, assuming that at each scale there is
a direct transfer of kinetic energy to the normal component. In
this refined approximation, the functional form of $E(k)$ becomes
slightly modified by the logarithmic correction, but the roll-off
exponent $-5/3$ remains unchanged. This analysis is closely
connected with problems of fully developed turbulence in classical
liquids, where similar logarithmic corrections have been proposed
by various authors. It should therefore be of interest to the
classical turbulence community, but details are beyond the scope
of this Letter.

The continuous approach for considering superfluid turbulence
based on eq.(2) is fully applicable in the limit
$\kappa\rightarrow 0 $. As pointed out by Volovik
\cite{GrishaJETP}, at finite $\kappa$ one has to ensure that, at
the smallest scale $r_{0}$, the "granularity" due to individual
vortices does not become important, so that the circulation
$v_{r0}r_{0}=q^2UR=q^2\kappa Re_{s}>\kappa$. This leads to an
important criterion $Re_{s}>1/q^2 \gg 1$. For small enough $q$,
when the turbulent cascade reaches small scales that would contain
only a few quantized vortices, the turbulent cascade will most
likely continue, but the form of the energy spectrum around and
beyond the quantum scale \cite{LSPRE}, $\ell_{q}\approx
(\varepsilon/\kappa^3)^{-1/4}$ must depend explicitly  on $\kappa$
\cite{QScale}.

In order to apply an analysis based on eq.(2), we must bear in
mind that this coarse-grained equation sufficiently accurately
describes the superfluid velocity field on the scale over which
the averaging is done. This approach  cannot therefore include
initial conditions similar to those commonly believed to apply in
counterflow turbulence in He II if only a single scale is assumed.
Such a distribution of vortices will most likely decay according
to the Vinen \cite{VinenOld} equation. It is well known and agrees
with simulations by Schwarz \cite{Schwarz} that there is a
critical self-sustaining counterflow velocity, above which the
turbulence is in dynamical equilibrium. Let us call it turbulent
state I, in accord with Tough's classification scheme
\cite{Tough}. According to computer simulations and a common
belief based on  the experiments of Awschalom {\it{et al}}
\cite{AMS} this state is, at least approximately, homogeneous. If
it contains just one scale, the vortex line density, $L$, ought to
decay as $1/t$ as follows from the Vinen equation and, according
to some experiments \cite{VinenOld, Milliken}, it does. Now let
increase the mean flow (i.e., the counterflow velocity $U_{cf}$),
assuming the normal fluid velocity profile remains flat, and
continue discussion in the reference frame where the normal fluid
is at rest. It is an established experimental fact that the
transition to state II occurs \cite{Tough, Ladner, noise}, with
distinctly different features. It has been a long lasting
challenge to explain the nature of this transition. We believe
that the answer is hidden in Volovik's analysis \cite{GrishaJETP}.
As he shows, there is a crossover between what he calls the
Kolmogorov and Vinen states of superfluid turbulence when
\begin{equation}
Re_{s}q^2=U_{cf}Rq^2/\kappa \simeq 1
\end{equation}
For higher counterflow velocities an analysis based on equation
(2) is valid and therefore a range of scales between the outer
scale, $R$ and a scale on which the circulation is of order
$\kappa$ (the quantum scale) occurs. Well above this transition,
(i.e., for high enough $Re_{s}$) there will be eddies spanning a
wide range of scales and within all these scales the counterflow
turbulence in the superfluid is of the Kolmogorov type.

But now that we have large superfluid eddies up to the size of the
channel, these will interact via mutual friction with the normal
fluid. Therefore the normal fluid is most likely driven into a
turbulent state, too, so that the high $Re_{s}$ counterflow
turbulence should be similar in character to the grid generated
turbulence. It is especially true for the case when the heater is
turned off and the turbulence decays. We believe that this is the
reason why an anomalous decay of counterflow turbulence in He II
was observed in the pioneering work of Vinen \cite{VinenOld} and
later by Schwarz and Rosen \cite{SR}. In our own decay experiments
\cite{Skr1}, we have observed that the temperature gradient along
the counterflow channel decays very fast when the heater is
switched off, so the flow can be considered as isothermal when the
second sound decay measurement is being performed. We therefore
expect that the decays of high $Re_{s}$ counterflow turbulence and
grid generated He II turbulence ought to display the same
character. And, indeed, it was clearly shown in experiments, that
for both towed grid generated He II turbulence \cite{Skr} and high
$Re_{s}$ counterflow turbulence \cite{Skr1}, most of the decay of
the vortex line density displays the same $t^{-3/2}$ power law.
This decay law follows from the spectral decay model \cite{Skr}
and is based on the existence of the Kolmogorov form of the 3d
energy spectrum, directly shown to be present in classically
generated turbulence both above and below the lambda point by
Maurer and Tabeling \cite{MT}.

That the normal fluid is turbulent in state II of the counterflow
He II turbulence is independently supported by the stability
analysis of Melotte and Barenghi  \cite{BM}.

The crossover to superfluid turbulence state II has been observed
in channels of circular and square crossection, but not in
channels of high aspect ratio rectangular crossection \cite{Hen}.
Naturally, the transition cannot take place if the size of the
sample intervenes. If some dimension of the channel is too small,
its physical size $R_{0}$ limits the size of possible eddies.

It has been stated many times (see, e.g., \cite{QTReview}) that
counterflow turbulence does not have any classical analog. On the
other hand, a close similarity between counterflow turbulence and
turbulent thermal convection in the heat transport efficiency has
been pointed out \cite{Skr1}. It seems now that both views are
justified. Counterflow turbulence in state I exists owing to the
finite value of the circulation quantum and therefore cannot have
any classical analog. However, the counterflow turbulence in state
II is closely similar to turbulent thermal convection in the same
way as the grid generated classical and He II turbulence closely
resemble each other \cite{Skr, Vinen}.

There are many experimental data that can be used in order to
verify, at least qualitatively, the phase diagram (see Fig.1)
suggested by Volovik \cite{GrishaJETP}. The recent experiment of
Finne {\it {et al}} \cite{N} provides evidence for a velocity -
independent transition from a laminar to a turbulent flow regime
in rotating $^3$He-B, where values of $q$ of order unity are
experimentally easily accessible. In He II these large values of
$q$ occur very close to the lambda point, where, to our knowledge,
no reliable measurements  exist that can be considered in the
frame of reference of the normal fluid. On the other hand, there
is ample experimental data on counterflow He II turbulence at
lower temperatures. However, the data on the transition into
superfluid state I (Vinen state) in tubes and capillaries of
various sizes cannot be reliably used here, as it is believed that
below this threshold the viscous normal fluid possesses a velocity
profile similar to a flow of ordinary viscous flow in a pipe. A
unique frame of reference in not, therefore, provided by the
normal fluid. However, Baehr {\it{et al}} \cite{ToughPipe} studied
the transition from dissipationless superflow to homogeneous
superfluid turbulence, when both ends of the pipe were blocked by
superleaks and the normal fluid inside the pipe thus remains
stationary, thereby providing this unique frame of reference.
These data, spanning the temperature range $1.3$~K~$<T<1.9$~K,
mark the transition from laminar flow into the Vinen state (state
I) and are shown in Fig. 1.

Various counterflow experiments clearly display the transition
from state I (Vinen) into state II (Kolmogorov) - the signature is
pronounced on temperature and pressure difference versus heat
input dependencies. We use here the data of Ladner, Childers and
Tough \cite{Ladner} (their Table I), assuming that in state I the
normal fluid profile is flat, again providing the unique frame of
reference with the normal fluid at rest. Let us point out that
this transition into a different flow regime is accompanied with a
pronounced increase of fluctuations \cite{noise}, characteristic
of phase transitions. The data \cite{Ladner} also clearly show
that on increasing the temperature the difference in counterflow
velocity between state I and II transitions decreases and around
$2$~K they become indistinguishable.

\begin{figure}[t]
\centerline{\includegraphics[width=1\linewidth]{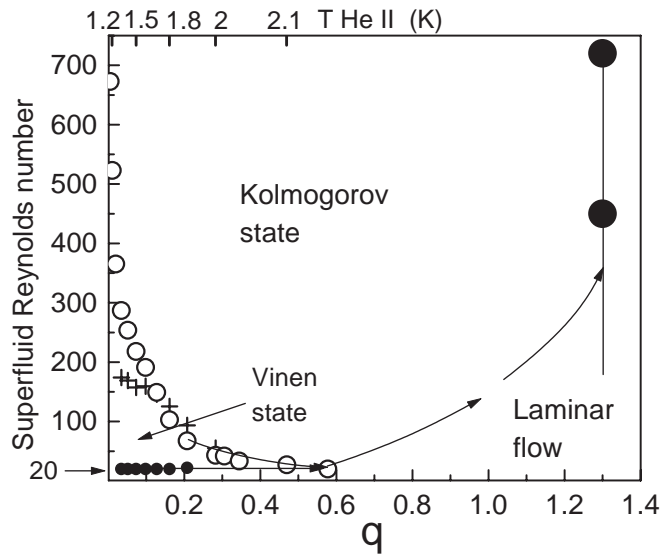}}
\medskip
\caption{The observed flow diagram of He II and $^3$He -B
superfluids in the unique frame of reference where the normal
fluid is at rest. The abscissa, $q$, represents the inverse of the
Reynolds number for superflow (for convenience the corresponding
temperature in He II is indicated on the upper axis), while the
ordinate, $Re_{s}$, represents the strength of circulation at the
outer scale of the flow in units of $\kappa$. The small filled
circles represent the onset of turbulent state I in pure superflow
of He II when the motion of the normal fluid was inhibited by
superleaks \cite{ToughPipe}; the crosses \cite{Ladner} and open
circles \cite{Chase} mark the transition from state I into state
II for counterflowing He II. The two big filled points mark
approximately the region where the onset of superfluid turbulence
has been observed by various methods of vortex loop injection into
rotating $^3$He-B in the vortex-free Landau state (for $\Delta T
\approx 0.05 ~T_{c}$ around $0.6 ~T/T_{c}$ at $29$~bar, see Fig.3
in \cite{N})}\label{HeII}
\end{figure}

As another set of experimental data marking the state I -state II
transition we have used the thermal conduction measurements of He
II in tubes of various diameters of Chase \cite{Chase}. We have
scanned the available experimental data and show in Fig. 2 that
they collapse onto a single curve if Reynolds number scaling is
applied. The open circles in Fig. 1 correspond to the onset of
state II.

We emphasize that the procedure used to acquire the data points
shown in the flow diagram is probably not very accurate for
several reasons, such as different temperature scales or
uncertainty in values of $q$, and more work is needed to map it
out accurately. We believe, however, that the essential physics is
displayed clearly and that Fig.1 strongly supports the ideas
underlying the physical problem of superfluid turbulence.

Let us comment here on the apparent disagreement between this
phase diagram and the computer simulations of Araki {\it {et al}}
\cite{Tsu} done in the zero temperature limit in the sense that
there is no normal fluid, strongly indicating the presence of
Kolmogorov scaling. This computer simulation introduces a cutoff
at the scale of the grid used for simulations, in that all vortex
rings or loops smaller than this size are removed from the flow.
This effectively introduces an artificial dissipative mechanism at
a prescribed length scale. We believe that any dissipative
mechanism acting at some small length scale (such as a Kelvin wave
cascade with subsequent phonon emission \cite{Vinen}) leads to a
Kolmogorov cascade in the continuum approximation, as the
assumptions for it are only that there is a range of scales where
dissipation is unimportant, that the form of the energy spectrum
only depends on $k$, and that the total energy decay rate
$\varepsilon=-dE/dt$ is independent of $k$. Dimensional analysis
then leads to the energy spectrum of the Kolmogorov form. The
physical mechanism of the dissipation is unimportant, so long as
it acts only on small scales.

In practice, there should be a crossover from the mutual friction
dissipation regime correctly described by Volovik's analysis into
a different one, probably based on vortex wave irradiation. The
governing equation (2) will have to be altered accordingly, and
similar analysis as in \cite{GrishaJETP} ought to be repeated. A
Kolmogorov cascade will most likely emerge again, as soon as the
smallest scale obtained with this new dissipation mechanism is
larger than the quantum scale.
\begin{figure}[t]
\centerline{\includegraphics[width=0.95\linewidth]{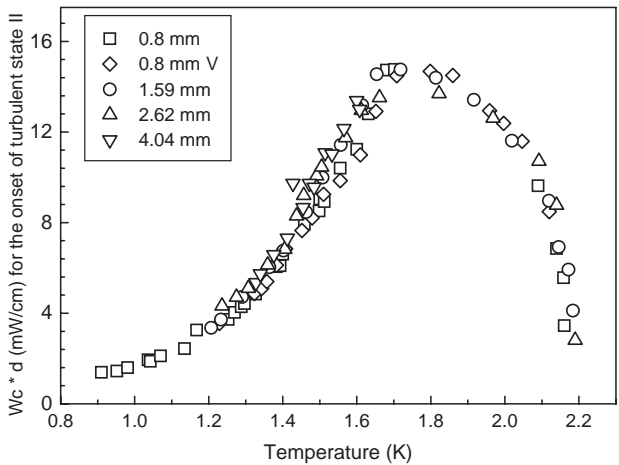}}
\medskip
\caption{Product of the critical heat input per unit area, $Wc$,
times the inner diameter of the used tube, $d$, marking the onset
of the turbulent state II versus temperature \cite{Chase} (two
different experimental methods have been used for a tube with
$d=0.8$~mm). The data obtained with tubes of various $d$ as
indicated collapse onto a single curve. Assuming that
$U_{cf}\propto Wc$,
the onset of turbulent state II occurs at any temperature at a
critical value of $Re_{s}$, shown in Fig.1. }\label{Chase}
\end{figure}

To conclude, we show that the extraordinary fluid properties of
quantum fluids give rise to the flow diagram suggested by Volovik
\cite{GrishaJETP}, containing two distinctly different turbulent
flow regimes called Vinen and Kolmogorov regimes. These can most
likely  be identified with the puzzling turbulent states I and II
according to the classification scheme of Tough \cite{Tough}.

Discussions with many colleagues, especially with P.V.E.
McClintock, M. Krusius, W.F. Vinen and G.E. Volovik are warmly
acknowledged. This research was supported by the Czech Grant
Agency, \#
202/02/0251.


\begin{thebibliography}{99}

\bibitem{Tough} J.T.\ Tough, Superfluid turbulence, in {\it Prog. in Low
Temp. Phys}. Vol.VIII, North-Holland, Amsterdam, (1982).
\bibitem{isothermal} This cannot be strictly true, as
dissipation in flowing normal fluid possessing finite viscosity
leads to heating in places of high vorticity and to a counterflow.
\bibitem{Awschalom} D.D. Awschalom, and K.W. Schwarz, Phys. Rev. Lett.
\textbf{52}, 49 (1984).
\bibitem{Sonin} E.B. Sonin, Rev. Mod. Phys. {\bf 59}, 87, (1987).
\bibitem{GrishaJETP} G.E. Volovik,
JETP Letters {\bf 78}, 553, (2003).
\bibitem{N} A.P. Finne {\it{et al.}}, Nature \textbf{424},1022 (2003) .
\bibitem{Russ} R. J. Donnelly and C. F. Barenghi, J. Phys. Chem. Data \textbf{27}
(1998) 1217.
\bibitem{Bevan} T.D.C. Bevan {\it et al.},
J. Low Temp. Phys. {\bf 109}, 423 (1997).
\bibitem{GrishaNote} G.E. Volovik,
private communication, cond-mat/0402035
\bibitem{LSPRE} L. Skrbek, J.J. Niemela, and K.R. Sreenivasan, Phys. Rev. \textbf{E64}, 067301 (2001).
\bibitem{QScale}
The exact functional form of the spectral energy density,
$\Phi(\varepsilon, k, \kappa)$, around and beyond $\ell_{q}$
cannot be written explicitly based on the dimensional analysis
similar to that of Kolmogorov (see \cite{LSPRE}), as it can
contain, in principle, any function of the dimensionless
combination $\varepsilon\kappa^{-3} k^{-4}$. However, the form of
$\Phi(\varepsilon, k, \kappa)$ can be judged if one uses
experimental data on the late decay of the grid generated
turbulence \cite{Skr}, relevant to scales of order
$\ell_{diss}\approx (\varepsilon/\nu_{n}^3)^{-1/4}$, where the
normal fluid can be considered at rest.
\bibitem{VinenOld} H.E. Hall, and W.F. Vinen, Proc. Roy. Soc. \textbf{A238}
204,205 (1956); W.F. Vinen, Proc. Roy. Soc. \textbf{A240} 114, 128
(1957); \textbf{A242} 489 (1957).
\bibitem{Schwarz} K.W. Schwarz, Phys. Rev. B \textbf{38}, 2398 (1988).
\bibitem{AMS} D.D. Awschalom, F.P. Milliken, and K.W. Schwarz,
Phys. Rev. Lett. \textbf{55}, 1372 (1984).
\bibitem{Milliken} F.P. Milliken, and K.W. Schwarz, Phys. Rev. Lett.
\textbf{48}, 1204 (1982).
\bibitem{Ladner} D.R. Ladner, R.K. Childers, and J.T. Tough,
Phys. Rev. \textbf{B13}, 2918 (1976).
\bibitem{noise} C.P. Lorenson, D. Griswold, V.U. Nayak, and J.T. Tough,
Phys. Rev. Lett. \textbf{55}, 1494 (1985).
\bibitem{Vinen} W.F. Vinen, Phys. Rev. \textbf{B61}, 1410 (2000).
\bibitem{SR} K.W. Schwarz, and J.R. Rosen, Phys. Rev. Lett.
\textbf{66}, 1898 (1991); Phys. Rev. \textbf{B44}, 7563 (1991).
\bibitem{Skr1} L. Skrbek, A.V. Gordeev, and F. Soukup,
Phys. Rev. \textbf{E67}, 047302 (2003).
\bibitem{Skr} L. Skrbek, J.J. Niemela, and R.J. Donnelly,
Phys. Rev. Lett. \textbf{85}, 2973 (2000).
\bibitem{MT} J. Maurer, P. Tabeling, Europhys. Lett. \textbf{43}, 29 (1998).
\bibitem{BM} D.J. Melotte, C.F. Barenghi, Phys. Rev. Lett.
\textbf{80}, 4181 (1998).
\bibitem{Hen} C.P. Lorenson, D. Griswold, V.U. Nayak, and J.T. Tough,
Phys. Rev.\textbf{B23}, 1494 (1985).
\bibitem{QTReview} W.F. Vinen, and J.J. Niemela,  J. Low Temp. Phys. \textbf{128}, 167 (2002).
\bibitem{ToughPipe} M.L. Baehr, L.B. Opatowsky, and J.T. Tough,
Phys. Rev. Lett. \textbf{51}, 2295 (1983).
\bibitem{Chase} C.E. Chase, Phys. Rev.\textbf{127}, 361 (1962);\textbf{131}, 1898 (1963).
\bibitem{Tsu} T. Araki, M. Tsubota, and S.K. Nemirovskii, Phys. Rev. Lett.  \textbf{89} (2002) 145301.




\end{thebibliography}
\end{document}